\documentclass[twocolumn,prd,showpacs,amsmath,amssymb]{revtex4}
\usepackage{graphicx,color}
\begin{document}
\title{Quantum uncertainty in distance measurement:\\ Holography and black hole thermodynamics}

\author{Michael~Maziashvili}
\email{maziashvili@hepi.edu.ge}\affiliation{Department of Physics,
Tbilisi
State University, 3 Chavchavadze Ave., Tbilisi 0128, Georgia \\
Institute of High Energy Physics and Informatization, 9 University
Str., Tbilisi 0186, Georgia }

\begin{abstract}
Exact conditions on the clock parameters corresponding to the
minimal uncertainty in distance measurement are derived in uniform
manner for any number of space time dimensions. The result
espouses the holography principle no matter what the number of
space time dimensions is. In this context the ADD braneworld model
is considered. Some remarks are made on deviation of holography as
well as of special relativity at the scales provided by the
cosmological constant. We also comment on the potential influence
of the background radiation on the uncertainty in length
measurement. The presence of unavoidable quantum uncertainty in
length measurement results in fluctuations of the black hole
thermodynamics that can be interested to address the information
loss problem. The quantum corrections to the black hole entropy
obtained in various scenarios are imperceptible because of these
fluctuations. At the Planck scale the fluctuations destroy the
thermodynamic picture of the black hole.

\end{abstract}

\pacs{03.65.-w, 03.65.Ta, 04.70.-s, 04.70.Dy, 4.50.+h}


\maketitle

The general principles of quantum mechanics with those of general
relativity where combined in papers \cite{SW} to study the problem
of quantum measurements of space-time distances. The key point is
that in ordinary quantum mechanics the space-time is essentially
absolute while in general relativity coordinates have no meaning
independent of observation. The limitations imposed by the quantum
mechanics on a distance measurement between events in space-time
were subsequently studied in \cite{NGD}. The inherent quantum
uncertainty in measuring the distance is shown to be proportional
to the one-third power of the length itself. This result is in a
good agreement with the holographic principle \cite{tHS} claiming
that the number of degrees of freedom contained in a spatial
volume scales as  an area of the surface enclosing this volume.
The authors \cite{SC} generalized the discussion proposed in
\cite{NGD} to a higher dimensional case and arrived at a
conclusion that the relation between length's uncertainty and the
length does not espouses the holographic principle. First we
demonstrate in a rather simple and transparent way that the
relation between length's uncertainty and the length does not
contradict the holographic principle.

Let us critically follow the consideration of this problem
proposed in \cite{NGD, SC}. In what follows we assume
$\hbar=c=k_B=1$. Our measuring device is composed of a spherical
clock localized in the region $\delta x=2r_c$ and mass $m$, which
also serves as a light emitter and receiver and a mirror that is
situated at $r$, ($r_c$ denotes the radius of the clock). To
simplify our consideration let us assume that the mirror does not
disturb the Schwarzschild space-time of the clock. So, we are
measuring a distance $l$ by sending a light signal to the mirror.
To be more precise let us assume that the measurement implies the
signal to start at $r_c$ and come back at the same point after the
reflection. The problem to be well posed we require $l\gg r_c
> r_g$. Assuming the minimal uncertainty the clock's spread in velocity
can be found as $ \delta v = \delta p/m =1/2m\delta x$. After the
time $t = 2l$ elapsed by light to travel along the closed path
clock-mirror-clock the total uncertainty during the measurement
takes the form
\[
\delta l = \delta x + {t\over 2m\delta x}~,\] which after
minimization with respect to $\delta x$ gives
\begin{equation}\label{minim}r_c\sim\sqrt{{l \over m}}~,~~~\Rightarrow~~~\delta l_{min}\sim\sqrt{{l
\over m}}~.\end{equation} The distance from $r_c$ to the position
of mirror $r$ measured by the light signal is given by

\[l=\int\limits_{r_c}\limits^{r}{x^{1+n}dx\over
x^{1+n}-Al_{p}^{2+n}m}~,\] where $n$ denotes the number of extra
dimensions and
\[A={8\Gamma\left({3+n\over 2}\right)\over (2+n)\pi^{{n+1\over
2}}}~.\] For any given values of $r_c,~m$ one can always choose
$r$ such that $l$ to have some concrete value. In papers
\cite{NGD, SC} the quantity $l-r+r_c$ as error caused by the
curvature is added to the $\delta l_{min}$ in Eq.(\ref{minim}) and
the resulting expression is minimized with respect to $m$. This
step is absolutely unclear, less motivated and unnecessary at all.
We emphasize once again that though the length depends explicitly
on clock's parameters, in Eq.(\ref{minim}) one can suppose some
fixed value of $l$ with no loss of generality, keeping in mind
that the $r$ can be assumed to be appropriately chosen for given
values of $r_c,~m$. In terms of Schwarzschild radius $r_s\sim
\left(l_{p}^{2+n}m\right)^{{1\over 1+n}}$ from Eq.(\ref{minim})
one simply gets
\[ \left(l_{p}^{2+n}m\right)^{{1\over 1+n}}\left({r_c\over
r_s}\right)\sim \sqrt{{l\over m}}~,~~\delta l_{min}\sim
\left(l_{p}^{2+n}m\right)^{{1\over 1+n}}\left({r_c\over
r_s}\right)~,\] which results in the following relation
\begin{equation}\label{lengunc}\delta l_{min}\sim \left(ll_p^{2+n}\right)^{{1\over
3+n}}\left({r_c\over r_s}\right)^{{1+n\over 3+n}}~.\end{equation}
From Eq.(\ref{lengunc}) one sees that the minimization of
uncertainty implies the size of the clock to be comparable to its
Schwarzschild radius $r_c/r_s\sim 1$. So the two relations given
by Eq.(\ref{minim}) together with the condition $r_c/r_s\sim 1$
give the minimal uncertainty
\begin{equation}\label{minlengunc}\delta l_{min}\sim
\left(ll_p^{2+n}\right)^{{1\over 3+n}}~,\end{equation} which is
consistent with the holographic principle. Correspondingly the
fluctuations of metric for a measurement in a space-time region
with the linear size $l$ are \[\delta g_{\mu\nu}\sim
\left({l_p\over l}\right)^{{2+n\over 3+n}}~.\] The
Eq.(\ref{minlengunc}) was previously derived in a bit different
way in \cite{Ng}. From the above discussion one simply concludes
that to perform the measurement of distance $l$ with minimal
uncertainty the mass of the clock has to be
\begin{equation}\label{clmass}m\sim l^{{1+n\over
3+n}}l_p^{-{4+2n\over 3+n}}~.\end{equation} An interesting point
that comes from the above consideration is that the minimal
inaccuracy in the distance measurement is determined by the
quantum uncertainty and the near horizon structure (behavior) of
gravity. So that the above result can not distinguish between
different gravity theories with the same expression for
gravitational radius.

Now let us look what happens in the case of brane. In what follows
we restrict ourselves to the ADD model \cite{ADD}. Without going
into much details let us merely recall a few basic features
relevant for our consideration. There is a low fundamental scale
of the order of $\sim$TeV, the standard model particles are
localized on the brane while the gravity is allowed to propagate
throughout the higher dimensional space. There is a length scale
$L$, size of extra dimensions, beneath of which the gravitational
interaction has the higher dimensional form whereas beyond this
scale we have the standard four-dimensional law. Postulating the
TeV fundamental scale one finds $L\sim 10^{30/n-17}$ \cite{ADD}.
The above discussion tells us that if brane localized observer
uses a clock with $r_s < L$ the uncertainty in length measurement
behaves as in Eq.(\ref{minlengunc}) while for $r_s > L$ one gets
\begin{equation}\label{minlenguncfour}\delta l_{min}\sim
\left(ll_p^2\right)^{1/ 3}~.\end{equation} Using the
Eq.(\ref{clmass}) one gets \[r_s\sim l^{{1\over
3+n}}l_p^{{2+n\over 3+n}}~.\] Correspondingly the number of
degrees of freedom contained in the region the linear size of
which is less than the characteristic scale \[h \sim
L^{3+n}l_p^{-(2+n)}~,\] scales as $\sim l^{2+n}$ and therefore is
not consistent with the holographic principle whereas for the
region with linear size grater than $h$ the holographic principle
on the brane is satisfied. For $n=2$ one finds a huge number for
the characteristic scale $h\sim 10^{54}$cm. Since in
four-dimensional space-time the holographic principle is
controlled by the Planck length, in that it holds if the linear
size of the region much exceeds the Planck size, one concludes
that the natural transition to the brane-world model from the
standpoint of holographic principle implies $h\sim l_p
~\Rightarrow ~L\sim l_p~\Rightarrow ~n\sim 30$.

Let us turn to the four-dimensional case and consider how the
above discussion can be modified at very large distances due to
presence of cosmological constant. Following the simple and
transparent arguments proposed in \cite{AS} the standard Heisenber
uncertainty relation gets modified due to gravitational
interaction leading therefore to the generalized uncertainty
principle (GUP). Namely, during the measurement the photon due to
gravitational interaction imparts to electron the acceleration
given by
\[a={l_p^2\,\delta E\over r^2}-\Lambda r~,\] where the cosmological
constant is supposed to be $\Lambda\sim 10^{-56}$cm$^{-2}$. By
taking into account that the characteristic time and length scale
for the interaction when one uses the energy $\Delta E$ for the
measurement is given by $\Delta E^{-1}$ \cite{LL} one gets for the
gravitational displacement of a particle being observed
\[\delta x_g\sim l_p^2\,\delta E-\Lambda\delta E^{-3}~.\]
By adding it to the position uncertainty one gets the the
following expression for the GUP \begin{equation}\label{gup}\delta
x\delta E \geq {1\over 2}+\alpha \left|l_p^2\,\delta
E^2-\Lambda\delta E^{-2}\right|~,\end{equation} where the
parameter $\alpha$ is of order unity. (Notice that the troublesome
aspect of this approach has to do with the use of classical
gravity at the Planck scale \cite{Ma}. The above discussion can
immediately be generalized to higher dimensional case as well
\cite{MMN}). So that due to gravitational interaction the
Heisenberg uncertainty relation is replaced by
\begin{equation}\label{ngup}\delta x\delta p \geq {1\over 2}+\alpha
\left|l_p^2\,\delta p^2-\Lambda\delta
p^{-2}\right|~.\end{equation} From this relation one sees that the
asymptotic regimes of quantum mechanics are determined by the
gravitational effects. To carry out the corresponding
phenomenological results one can consider different quantum
mechanical problems via the Hilbert space representation of this
uncertainty relation. The above discussion tells us that the size
of the clock giving the minimal uncertainty scales as
$(l_p^2l)^{1/3}$ and can be arbitrarily large by letting $l$ to
grow correspondingly. From Eq.(\ref{ngup}) one sees that the GUP
in the limit $\delta p\ll \Lambda^{1/2}$ (corresponding to $\delta
x\gg \Lambda^{-1/2}$) is dominated by the last term. So assuming
the size of the clock is large enough one gets $\delta p^3\sim
\Lambda /l_p^2m$ and correspondingly
\[\delta l_{min}\sim (\Lambda^{1/3}l_p^2l)^{3/7}~.\] So that the
number of degrees scales as \[{l^3\over \delta l^3_{min}}\sim
{l^{12/7}\over (\Lambda^{1/3}l_p^2)^{9/7}}~,\] and thereby does
not agree with the holographic principle. Combining together one
finds that the holographic principle based on the minimal quantum
uncertainty in length measurement is valid in the region
\[l_p\ll l\ll \Lambda^{-3/2}l_p^{-2}~.\] Since the scale $\Lambda^{-3/2}l_p^{-2}\sim
10^{150}$cm is extremely large this upper bound may be in fact of
theoretical interest only.

Let us consider the situation when the clock is embedded into the
background radiation with a temperature $T$. Because of the
background temperature the clock has a mean velocity
$\sim\sqrt{T/m}$. Correspondingly, the Eq.(\ref{minim}) gets
modified
\[r_c\sim\sqrt{{l \over m}}~,~~~~~~\delta l_{min}\sim\sqrt{{l
\over m}}+l\sqrt{{T\over m}}~.\] Assuming the gravitational radius
of the clock is not changed significantly due to background
radiation and repeating the above discussion (by imposing the
minimization condition $r_c\sim r_s$) one finds
\begin{equation}\label{bacminuncer}\delta l_{min}\sim
(l_p^2l)^{1/3}+\sqrt{T}\,l_p^{2/3}l^{5/6}~.\end{equation} The
fluctuations of metric in the region with linear size $l$ can be
evaluated as \[\delta g_{\mu\nu}\sim \left({l_p\over
l}\right)^{2/3}+\sqrt{T}\,{l_p^{2/3}\over l^{1/6}}~.\] If
$\sqrt{l}\gg 1/\sqrt{T}$ the latter term in Eq.(\ref{bacminuncer})
becomes dominant and correspondingly
\[{l^3\over \delta l_{min}^3}\sim {l^{1/2}\over T^{3/2}l_p^2}~.\]
So that the holographic principe holds for $\sqrt{l}\ll
1/\sqrt{T}$. By taking $T\approx 2.7$K one finds that $1/\sqrt{T}
\sim \sqrt{\mbox{cm}}$. So that the upper bound on the holographic
principle becomes quite small in this case. For this value of
temperature from Eq.(\ref{bacminuncer}) one gets that uncertainty
for $l\sim 10^{28}$cm is about $\delta l_{min}\sim 30$cm.

Discussing in the spirit of Doubly Special Relativity \cite{AC}
from Eq.(\ref{gup}) one concludes that the presence of $\Lambda$
makes the Doubly Special Relativity more special. Namely, omitting
the gravitational displacement from Eq.(\ref{gup}) one simply
recognizes a well known result in quantum field theory \cite{BLP}
that the minimal localization width of the quantum in its rest
frame is controlled by the Compton wavelength. But the covariance
of Eq.(\ref{gup}) requires the deformation of special relativity
at the Planck scale \cite{AC} as well as at the scale $\delta
E\lesssim \Lambda^{1/2}$ (corresponding to the length scale
$\gtrsim \Lambda^{-1/2}\sim 10^{28}$cm).

Finally let us consider the influence of the unavoidable quantum
uncertainty in length measurement on the black hole physics.
Hawking's discovery of black hole radiance established a deep
connection between gravitation, quantum theory and thermodynamics
\cite{Ha}. That the black holes may be the thermodynamic objects
with a well defined temperature and entropy was conjectured by
Bekenstein \cite{Be}. The temperature of the black hole radiation
is given by the surface gravity at the horizon
\[T={f'(r_g)\over 4\pi}~,\] where $f(r)=1-2l_p^2\,m/r$, and the entropy is expressed
through the area of the event horizon $A=4\pi r_g^2$, \[S={A\over
4l_p^2}~.\] Quantum corrected black hole entropy derived in
different scenarios \cite{RSVSABCK} and multitudinous subsequent
papers has the form

\begin{equation}\label{qcentropy}S={A\over 4l_p^2}+\beta\ln{A\over l_p^2}+O\left({l_p^2\over
A}\right)~,\end{equation} where the model dependent parameter
$\beta$ is of order unity. However, we want to notice that the
above defined thermodynamic quantities are given with an inherent
uncertainty due to intrinsic inaccuracy of the horizon. Namely,
since one can never know the distance $l$ to a better accuracy
than $\sim (l_p^2\,l)^{1/3}$ the gravitational radius also maybe
known with the accuracy $\delta r_g\sim(l_p^2r_g)^{1/3}$. This
uncertainty of the horizon induces the fluctuations of the above
mentioned thermodynamic parameters of the black hole
\begin{eqnarray}\label{fluct}\delta T\sim {1\over r_g-\delta r_g}-{1\over r_g}~,~~\delta S\sim {(r_g+\delta r_g)^2-r_g^2\over l_p^2}~.\end{eqnarray}
Assuming $\delta r_g\ll r_g$ one gets
\[\delta T\sim {\delta r_g\over r_g^2}\sim
{l_p^{2/3}\over r_g^{5/3}}~,~~\delta S\sim r_g\delta r_g\sim
\left({r_g\over l_p}\right)^{4/3}~.\] The smallness of
fluctuations in comparison with the thermodynamic quantities
requires $(r_g/l_p)^{2/3}\gg 1$ which is nothing but the condition
$\delta r_g\ll r_g$. By taking into account that
$(r_g/l_p)^{4/3}>\ln(r_g/l_p)$ for $r_g/l_p>1$ and
$(r_g/l_p)^{4/3}\gg \ln(r_g/l_p)$ when $r_g/l_p \gg 1$ one simply
concludes that quantum corrections to the entropy
Eq.(\ref{qcentropy}) Can not be discernible because of
fluctuations.

As it was observed by Hawking \cite{Haw} the black hole radiation
described in \cite{Ha} raises a information loss puzzle for the
detailed form of the radiation does not depend on the detailed
structure of the body that collapsed to form the black hole.
Potentially the fluctuations of the radiation can hide the
information about the collapsing body. Namely, the fluctuations
inside the uncertainty band given by Eq.(\ref{fluct}) can be quite
intricate and correlated in a subtle way to store any information
about the initial quantum state. Information is lost in this case
in the sense that we are unable to keep track of these
fluctuations but one can no longer claim that the information is
destroyed as it is actually implied by the information loss
paradox. So that the presence of quantum uncertainty in black hole
thermodynamics mitigate the infirmation loss paradox.

From Eq.(\ref{fluct}) one sees that the fluctuations near the
Planck scale become of the order of thermodynamic quantities
themselves and therefore destroy the thermodynamical picture. So
that one should be cautious about the drawing sweeping conclusions
based on the black hole thermodynamics at the Planck scale. One
can not be sure what happens next when the black hole has
evaporated down to the Planck size without a deeper understanding
of Planck scale physics. On the other hand this restriction maybe
of principle character like the Heisenberg uncertainty principle.
Notice that the uncertainty of the entropy Eq.(\ref{fluct})
describes the inaccuracy of the holographic principle \cite{tHS}
in turn.

\vspace{0.4cm}

\begin{acknowledgments}
The author is indebted to M.~Gogberashvili, Z.~Kepuladze,
P.~Kurashvili and P.~Midodashvili for useful conversations as well
as Y.~J.~Ng for useful correspondence. The work was supported by
the \emph{INTAS Fellowship for Young Scientists}, the
\emph{Georgian President Fellowship for Young Scientists} and the
grant \emph{FEL. REG. $980767$}.
\end{acknowledgments}


\end{document}